\documentstyle[aps]{revtex}
\begin{document}
\draft
\preprint{}
\title{Field theoretic study of light hypernuclei}
\author{P.K. Panda}
\address{Institute of Physics, Bhubaneswar-751005, India.
\footnote[1]{Present address: Institute of
Astrophysics, NAPP Group, Bangalore-560034, India\\ 
email: prafulla@iiap.ernet.in}}
\author{R. Sahu}
\address{Physics Department, Berhampur University, Berhampur-760007, India.}
\date{\today}
\maketitle
\begin{abstract}
A nonperturbative field theoretic calculation has been made for the 
$s$-shell hypernuclei. Here we dress the $\Lambda$- and $\Sigma$- hypernuclei 
with off-mass shell pion pairs. The analysis replaces the scalar isoscalar 
potential by quantum coherent states. The binding energies of $^{4+n\Lambda}$He 
$(n=0,1,2)$ agree quite well with the RMF result of Greiner. The
experimental binding energies of $^4$He and $^3$He are reasonably well
reproduced in our calculation. A satisfactory description of the relevant 
experimental $\Lambda$- and $\Sigma$- separation energies has been obtained.
\end{abstract}


\section{Introduction}

The investigation of the possible extension of the periodic table in
the sector of strangeness 
\cite{grein} has been a topic of current interest.
Over the years, various theoretical calculations 
\cite{grein,rufa,dov1,scha1,dov2,dov3,jen,dalitz,bodm} have been carried out
to understand the properties of the hypernuclei with
varying success. Greiner and his group have performed the relativistic
mean field (RMF) approach to study hypernuclei. The RMF has been quite
successful to describe baryons interacting through mesons. This model has been
used to study the binding energy of the $\Lambda$-hypernuclei, 
$^{4+n\Lambda}$He, $^{16+n\Lambda}$O and $^{40+n\Lambda}$Ca. They found the
binding energy of these nuclei to increase when hyperons are added to
normal nuclei due to the opening of new degrees of freedom. They have also 
studied exotic multi-strange nuclei which are formed by adding
$\Sigma$'s and $\Xi$'s within the above RMF approach. Calculations
have also been performed using Monte-carlo method with the potential well
depth calculated variationally with fermi-hyperneted chain method \cite{bodm}.
However, they are unable to reproduce simultaneously the binding energies
of $^4$He and the neighbouring hypernuclei. Not many theoretical calculations 
exist with $\Sigma$-hyperons. Hence it would be quite intersting to 
develop a model which simultaneously reproduces
the binding energies of $^4$He and $^3$He
as well as the neighbouring hypernuclei (both $\Lambda$ and $\Sigma$
hypernuclei).

Recently we have developed a nonperturbative technique 
based on quantum field theory \cite{panda,panda1,mishra90} to study light 
nuclei and nuclear matter. Here the $\sigma$-meson  effects are 
simulated through isosinglet scalar cloud of pair of off-shell pions with a
coherent state. This corresponds to the quantum picture of classical fields of
Walecka model \cite{walecka}. Such a construct also includes
higher order effects \cite{panda,panda1,mishra90}. In this model the 
nucleus contains, with a small probability, a finite expectation value 
for off mass shell scalar isosinglet pion pairs which may have observable 
effects. The calculation is based on the nonperturbative techniques of field 
theory in which the nucleons are dressed with off mass-shell pions.

This model has been successfully applied to study the binding
energy and other properties of deuteron \cite{panda} and $^4$He 
\cite{panda1} and nuclear matter \cite{mishra90}. In view of the above
successes, it would be quite interesting to apply the same to a study 
of the light hypernuclei here.

In section II, we construct the effective Hamiltonian along with a 
scalar isoscalar coupling of pions with the nucleons and hyperons and 
construct translationally invariant states explicitly 
with baryons and off mass shell pion cloud.
In section III we calculate the energy
expectation value and extremise it. 
We discuss the results and possible experimental signatures in sectionn IV.
The conclusions drawn from the present study are given in section V.

\section{Theory}

The Langrangian density for the pion, nucleon and hyperon system is taken as
\cite{panda}
\begin{eqnarray}
{\cal L}&=&{\bar N}(i\gamma^\mu\partial_\mu 
- M_N+G_{\pi NN} \gamma_5\phi)N
+{\bar \Lambda}(i\gamma^\mu\partial_\mu - M_\Lambda
+G_{\pi\Lambda\Lambda}\gamma_5\phi)\Lambda\nonumber \\
&+& \frac{1}{2}(\partial_\mu\phi_i\partial^\mu\phi_i-\mu^2\phi_i\phi_i)
\label{eq1}
\end{eqnarray}
where $M_N$, $M_\Lambda$ and $\mu$ are the masses of nucleon,
hyperon and pion, respectively. $G_{\pi NN}$ and $G_{\pi\Lambda\Lambda}$
are the coupling constants for pion-nucleon and pion-hyperon.
We shall consider equation (\ref{eq1}) in the non-relativistic limit. 
Then the effective Hamiltonian density for the system with nucleons, 
hyperons and pions are given as \cite{panda,panda1}
\begin{equation}
{\cal H}({\bf z})={\cal H}_N({\bf z})+{\cal H}_\Lambda({\bf z})
+{\cal H}_M({\bf z}) + {\cal H}_I({\bf z}).
\label{eq2}
\end{equation}
Here  the nucleon Hamiltonian density ${\cal H}_N({\bf z})$ and hyperon 
Hamiltonian density ${\cal H}_\Lambda({\bf z})$ are given by
\begin{mathletters}
\begin{equation}
{\cal H}_N({\bf z})=N_\alpha
({\bf z})^\dagger(M_N-{{\bf \bigtriangledown}_z^2\over 2M_N})N_\alpha({\bf z}),
\label{eq3}
\end{equation}
\begin{equation}
{\cal H}_\Lambda({\bf z})= \Lambda_\beta
({\bf z})^\dagger(M_\Lambda-{{\bf \bigtriangledown}_z^2\over 2M_\Lambda})
\Lambda_\beta({\bf z}).
\label{eq4}
\end{equation}
The meson Hamiltonian density, ${\cal H}_M({\bf z})$, is given by
\begin{equation}
{\cal H}_M({\bf z})=\frac{1}{2}[(\partial_0\phi_i({\bf z}))^2+(\bigtriangledown
\phi_i({\bf z}))^2+\mu^2(\phi_i({\bf z}))^2]
\label{eq5}
\end{equation}
and the interaction Hamiltonian density between baryons and pions is given
by
\begin{eqnarray}
{\cal H}_I({\bf z}) &=& i{G_{\pi N N}\over 2M_N} N_\alpha({\bf z})^\dagger
({\bf \sigma} \cdot{\bf\bigtriangledown} \phi({\bf z})) 
N_\alpha({\bf z}) 
+{G^2_{\pi N N}\over 2M_N} N_\alpha({\bf z})
^{\dagger}N_{\alpha}({\bf z})\phi_{i}({\bf z})\phi_{i}({\bf z})\nonumber\\
&+& i{G_{\pi \Lambda \Lambda}\over 2M_\Lambda} \Lambda_\beta({\bf z})^\dagger
({\bf \sigma} \cdot{\bf\bigtriangledown} \phi({\bf z})) \Lambda_\beta({\bf z}) 
+{G^2_{\pi \Lambda \Lambda}\over 2M_\Lambda} \Lambda_\beta({\bf z})
^{\dagger}\Lambda_\beta({\bf z})\phi_{i}({\bf z})\phi_{i}({\bf z}).
\label{eq6}
\end{eqnarray}
\end{mathletters}
In addition we include repulsive term ${\cal H}_R({\bf z})$ 
and Coulomb term ${\cal H}_C({\bf z})$ in the Hamiltonian density 
which are discussed later. We note that here we
take the ordered products for the above expressions, so that
the vacuum energy is zero. In the above 
$\alpha$ stands for both isospin and spin indices, such that
$\alpha=1,2$ stands for proton with spin $=\pm 1/2$ and $\alpha=3,4$ 
stands for neutron with spin $=\pm 1/2$ and $\beta$ stands for hyperons.
Here $N_\alpha^\dagger$
and $N_{\alpha}$ are the nucleon creation and annihilation operator.
$\Lambda_\beta^\dagger$ and $\Lambda_\beta$ are the hyperon
creation and annihilation operators. The matrix $\phi$ is given as 
$\phi({\bf z})= \tau_i\phi_i({\bf z})$. We expand the field operator 
$\phi_i({\bf z})$ in terms of creation and annihilation operators of off-mass
shell mesons satisfying equal time algebra as \cite{misra87}
\begin{equation}
\phi_i({\bf z})={1\over\sqrt{2\omega_z}}(a_i({\bf z})^\dagger+a_i({\bf z})).
\label{eq7}
\end{equation}
In the perturbative basis we have
$\omega_z={(\mu^2-{\bf\bigtriangledown}_z^2)}^{1\over 2}$. We shall still use 
this, but we note that since we shall take an arbitrary number of pions in a 
coherent manner as given later, the results shall be nonperturbative.
Here $i=1,2,3$ stand for isospin indices of pions. $a_i({\bf z})$
and $a_i({\bf z})^\dagger$ are the annihilation and creation
operators of the mesons.

We shall now consider the system with $n_N$ nucleons and $n_\Lambda$ hyperons
with a dressing of meson pairs. For this purpose, we shall first define the
state in a manner which shall be more convenient for field theoretic 
calculations and then consider the energy expectation values to obtain
the nucleon wavefunction and the meson dressing. Clearly, the two pion
dressing \cite{panda,panda1} will simulate the effect of hypothetical
$\sigma$-meson exchange. One objective here is to show that such a
picture of nuclear structure with Walecka's model generalised to include 
quantum two pion condensate can be a viable alternative, and we need not
take separately a $\sigma$-meson. With this in mind, we start with the 
notation for the form of nucleon and hyperon creation operators as
\begin{mathletters}
\begin{equation}
N^c_\alpha({\bf x})^\dagger=\int U^N_\alpha({\bf x}-
{\bf x}_\alpha)N_\alpha({\bf x}_\alpha)^\dagger d{\bf x}_\alpha,
\label{eq8}
\end{equation}
\begin{equation}
\Lambda^c_\alpha({\bf x})^\dagger=\int U^\Lambda_\alpha({\bf x}-
{\bf x}_\alpha)\Lambda_\alpha({\bf x}_\alpha)^\dagger d{\bf x}_\alpha,
\label{eq9}
\end{equation}
\end{mathletters}
so that, the creation operator for the helium system is taken as
\begin{equation}
{\cal N}_S({\bf x})^\dagger=\prod_{\alpha=1}^{n_N}
N^c_\alpha({\bf x})^\dagger
\prod_{\beta=1}^{n_\Lambda}\Lambda^c_\beta({\bf x})^\dagger.
\label{eq10}
\end{equation}
The background meson fields are taken through a coherent state type 
of formalities with the meson cloud creation operator being \cite{panda,misra87}
\begin{equation}
{\cal M}({\bf x})^\dagger=e^{B({\bf x})^\dagger},
\label{eq11}
\end{equation}
where
\begin{equation}
B({\bf x})^\dagger={1\over 2}\int f_M\left({\bf x}-{{\bf z}_1+
{\bf z}_2\over 2}\right)f({\bf z}_1-{\bf z}_2)a_i({\bf z}_1)^\dagger
a_i({\bf z}_2)^\dagger d{\bf z}_1 d{\bf z}_2.
\label{eq12}
\end{equation}
In the above, we are taking the nucleus to be centered around the point 
${\bf x}$, and so also the mesons. The mesons are taken only as pairs,
as isospin singlets. In mean field approximation, we shall
have the above conststruct giving isoscalar meson pairs with
{\it even} parity. We aproximate the arbitrary number of 
meson pairs through the two functions, $f_M({\bf r})$ and
$f({\bf r})$, the first indicating the distribution of mesons
with respect to ${\bf x}$, and the second,the mutual correlation
of the mesons.

We assume that $U^N_\alpha({\bf r})$, $U^\Lambda_\alpha({\bf r})$ and 
$f_M({\bf r})$ are so normalised that 
\[\int |U^N_\alpha({\bf r})|^2 d{\bf r}
=\int |U^\Lambda_\alpha({\bf r})|^2 d{\bf r}
=\int |f_M({\bf r})|^2 d{\bf r}=1.\]
As earlier \cite{panda1}, we take
\begin{mathletters}
\begin{equation}
U^N_\alpha({\bf r})=(\pi R_N^2)^{-3/4}e^{-{r^2\over 2R_N^2}},
\label{eq13}
\end{equation}
\begin{equation}
U^\Lambda_\alpha({\bf r})=(\pi R_\Lambda^2)^{-3/4}e^{-{r^2\over 2R_\Lambda^2}},
\label{eq14}
\end{equation}
\begin{equation}
f_M({\bf r})=(\pi R_M^2)^{-3/4}e^{-{r^2\over 2R_M^2}},
\label{eq15}
\end{equation}
and
\begin{equation}
f({\bf r})=a(\pi R_\pi^2)^{-3/4}e^{-{r^2\over 2R_\pi^2}},
\label{eq16}
\end{equation}
\end{mathletters}
where $a$, $R_{M}$, $R_{\pi}$, $R_N$ and $R_\Lambda$ are arbitrary
parameters which are determined variationally.

With the above construction, we shall now define the state of the system 
with $n_N$ nucleons and $n_\Lambda$ hyperons located at ${\bf x}$ as
\begin{equation}
|^AHe({\bf x})>=N_R(\pi R_N^2)^{-3/4}
{\cal N}_S({\bf x})^\dagger{\cal M}({\bf x})^\dagger | vac>
\label{eq17}
\end{equation}
where $N_R$ is the normalization constant, $A=n_N+n_\Lambda$ is the total
number of baryons.  We construct translationally invariant state of momentum
${\bf p}$ as \cite{bolsterli76}
\begin{equation}
|^AHe({\bf p})>=N_{R}{(2\pi)}^{-3/2}{(\pi R_{N}^{2})}^{-3/4}
\int d{\bf x} e^{i{\bf p}.{\bf x}} {\cal N}_S({\bf x})^{\dagger}
{\cal M}({\bf x})^{\dagger}\mid vac>.
\label{eq18}
\end{equation}
The details regarding the evaluation of the normalisation constant are
given in Appendix A. The normalisation constant is given by
\begin{equation}
N_R^{-2}=S_0
\label{eq35}
\end{equation} 
with
\[ S_0=\sum_{n=0}^{\infty}\left({3a^{2}\over 2}\right)^n
{1\over n!}\left({1\over a_n}\right)^{3/2}\]
where
\begin{equation} 
a_n= {n_N\over 4}+{n_\Lambda R_N^2\over 4R^2_\Lambda}+{nR_N^2\over 4R_M^2}.
\label{eq36}
\end{equation}
Then we shall proceed to evaluate the energy expectation values for the given 
system {\it at rest}. 

\section{Energy expectation values and its extremization}
With ${\cal H}({\bf z})$ as in equation (\ref{eq2}) the energy operator is
\begin{equation}
H=\int {\cal H}({\bf z})d{\bf z}.
\label{eq37}
\end{equation}
We are thus to find the expressions corresponding to equations (\ref{eq2})
to obtain the energy. The advantage of the present approach is that with the
ansatz function for the pions, we include an arbitrary number of pion  pairs
in a coherent manner with equal time algebra, which makes the contributions
calculable while retaining basically nonperturbative `higher order' effects
without any truncation. Clearly, at present we are concentrating on the
relevant attractive channel corresponding to $\sigma$-meson earlier 
\cite{walecka} without using the above unphysical "particle". The picture is 
also completely quantum mechanical instead of classical.

\subsection{Nucleon and hyperon kinetic energy}
We shall now calculate the energy expectation values for the 
above configurations. We proceed to evaluate the expectation value 
of $H_N$. The expectation $h_N$ for the same is defined 
through the equation, for ${\bf p}=0$,
\begin{equation}
<^AHe({\bf p}')| H_N |^AHe({\bf p})>=h_N\delta({\bf p}'-{\bf p}).
\label{eq38}
\end{equation}
In the above, we ignore the mass term. Then substitution of equation 
(\ref{eq3}) and (\ref{eq18}) gives that
\begin{equation}
h_N=N_R^2{(\pi R_N^2)}^{-3/2}\int g_N^{K.E}({\bf r})g_M({\bf r})d{\bf r},
\label{eq39}
\end{equation}
where
\begin{eqnarray}
g_N^{K.E}({\bf x}'-{\bf x})&=&\sum_{\alpha=1}^{N_n}
\int U_\alpha^N({\bf x}'-{\bf x}_\alpha)^*\left(-
{{{\bf\bigtriangledown}_{x_\alpha}^2}\over 2M_N}\right)U^N_{\alpha}({\bf x}-
{\bf x}_\alpha)d{\bf x}_\alpha
\prod_{\beta\not=\alpha}
\rho^N_\beta({\bf x}'-{\bf x})
\prod_{\gamma=1}^{n_\Lambda} \rho^\Lambda_\gamma({\bf x}'-{\bf x})\nonumber\\ 
&=&\sum_{\alpha=1}^{n_N}\left(-
{{{\bf\bigtriangledown}_x^2}\over 2M_N}\right)\rho^N_\alpha({\bf x}'
-{\bf x})\prod_{\beta\not=\alpha}\rho^N_\beta ({\bf x}'-{\bf x})
\prod_{\gamma=1}^{n_\Lambda}\rho^\Lambda_\gamma({\bf x}'-{\bf x}).
\label{eq40}
\end{eqnarray}
Substituting equations(\ref{eq15}and (\ref{eq22}) we then obtain from 
equation (\ref{eq40}) as 
\begin{equation} 
g_N^{K.E}({\bf r})=\left[ {3\over M_NR_N^2}-{{\bf r}^2\over
2M_NR_N^4}\right]\exp{\left(-{n_N{\bf r}^2\over 4R_N^2}\right)}.
\label{eq41}
\end{equation}
Therefore equation (\ref{eq39}) becomes
\begin{equation} 
h_N={n_N\over 4}{3\over M_NR_N^2}\left(1-{S_1\over 4S_0}\right),
\label{eq42}
\end{equation} 
where 
\begin{equation} 
S_1=\sum_n{1\over n!}\left({3a^2\over 2}\right)^n
\left({1\over a_n}\right)^{5/2}.
\label{eq43}
\end{equation}
Similarly the kinetic energy for the hyperon is given as
\begin{equation} 
h_\Lambda={n_\Lambda\over 4}{3\over {M_\Lambda}R_{\Lambda}^2} 
\left(1-{S_1\over 4S_0}\right).
\label{eq44}
\end{equation} 

\subsection{Meson kinetic energy}
With the meson field operator expressions as in equation (\ref{eq7}) we
may write equation (\ref{eq5}) as
\begin{equation}
{\cal H}_M({\bf z})= a_i({\bf z})^\dagger\omega_z a_i({\bf z}).
\end{equation}
The meson kinetic energy can be calculated through the equation
\begin{equation}
<^AHe({\bf p}')| H_M | ^AHe({\bf p})>= h_M\delta({\bf p}'-{\bf p}).
\label{eq45}
\end{equation}
Proceeding exactly in the same way as in ref.\cite{panda1} we get
\begin{equation} 
h_M=\sum_{n}({3a^2\over 2})^n{1\over n!}h_M^{(n)},
\label{eq49}
\end{equation}
where 
\begin{equation} 
h_M^{(n)}={3A^2\over S_0}\;{\left({1\over \pi}
{{4R_M^2R_\pi^2}\over{a_n(4R_M^2+R_\pi^2)+R_N^2}}
\right)}^{3/2}\int\exp{\left[-{{4R_M^2R_\pi^2a_{n+1}}\over
{a_n(4R_M^2+R_\pi^2)+R_N^2}}{\bf q}^2\right]}\omega({\bf q})d{\bf q}.
\label{eq50}
\end{equation}

\subsection{Interaction energy}
We now calculate the interaction energy. For this purpose
we first note that
\begin{equation} 
:\phi_i({\bf z})\phi_i({\bf z}):=\phi_i^{cr}({\bf z})\phi_i^{cr}
({\bf z})+\phi_i^{an}({\bf z})\phi_i^{an}({\bf z})+2\phi_i^{cr}({\bf z})
\phi_i^{an}({\bf z}),
\label{eq51}
\end{equation}
where we have substituted e.g.
\begin{equation}
\phi_i^{cr}({\bf z})={1\over \sqrt{2\omega_z}}a_i({\bf z})^\dagger.
\label{eq52}
\end{equation}
The expectation $h_I$ for the same is defined through the equation,
for ${\bf p} =0$ as
\begin{equation}
<^AHe({\bf p}')| H_I | ^AHe({\bf p})>=h_I\delta({\bf p}'-{\bf p}).
\label{eq53}
\end{equation}
Substitution of equation (\ref{eq6}) and (\ref{eq18}) we have
\begin{eqnarray}
&&h_I\delta({\bf p}'-{\bf p})=\nonumber\\ && N_R^2{(\pi R_N^2)}^{-3/2}
\left[{G^2_{\pi N N}\over 2M_N}
\int g_N^{INT}({\bf x}'-{\bf z},{\bf x}-{\bf z})
g_M^{INT}({\bf x}'-{\bf z},{\bf x}-{\bf z})d{\bf z} 
e^{-i{\bf p}'.{\bf x}'+i{\bf p}.{\bf x}}d{\bf x}'d{\bf x} 
\right.\nonumber \\ & +&
\left.{G^2_{\pi \Lambda\Lambda}\over 2M_\Lambda}\int 
g_\Lambda^{INT}({\bf x}'-{\bf z},{\bf x}-{\bf z})
g_M^{INT}({\bf x}'-{\bf z},{\bf x}-{\bf z})d{\bf z} 
e^{-i{\bf p}'.{\bf x}'+i{\bf p}.{\bf x}}d{\bf x}'d{\bf x} \right]
\label{eq54}
\end{eqnarray}
where
\[g_N^{INT}({\bf x}'-{\bf z},{\bf x}-{\bf z})=<vac
| {\cal N_S}({\bf x}')N_\alpha({\bf z})^\dagger N_\alpha({\bf z})
{\cal N_S}({\bf x})^\dagger | vac>,\]
\[g_\Lambda^{INT}({\bf x}'-{\bf z},{\bf x}-{\bf z})=<vac
| {\cal N_S}({\bf x}')\Lambda_\alpha({\bf z})^\dagger \Lambda_\alpha({\bf z})
{\cal N_S}({\bf x})^\dagger | vac>\]and 
\[g_M^{INT}({\bf x}'-{\bf z},{\bf x}-{\bf z})=<vac |{\cal M}({\bf x}')
\phi_i({\bf z})\phi_i({\bf z}){\cal M}({\bf x})^\dagger | vac>\]
We first consider the cr-cr term of equation (\ref{eq51}). Noting that
\begin{eqnarray}
&&<vac | {\cal N_S}({\bf x}')N_\alpha({\bf z})^\dagger N_\alpha({\bf z})
{\cal N_S}({\bf x})^\dagger | vac>\nonumber\\ 
&=&n_N\exp\left(-{(n_N-1){\bf r}^2\over 4R_N^2}
-{n_\Lambda {\bf r}^2 \over 4R_{\Lambda}^2} \right)
U^N({\bf x}'-{\bf z})^*U^N({\bf x}-{\bf z}),
\label{eq55}
\end{eqnarray}
\begin{eqnarray}
&&<vac | {\cal N_S}({\bf x}')\Lambda_\beta({\bf z})^\dagger 
\Lambda_\beta({\bf z})
{\cal N_S}({\bf x})^\dagger | vac>\nonumber\\ 
&=&n_\Lambda\exp\left(-{(n_\Lambda-1){\bf r}^2\over 4R_\Lambda^2}
-{n_N {\bf r}^2 \over 4R_N^2} \right)
U^\Lambda({\bf x}'-{\bf z})^*U^\Lambda({\bf x}-{\bf z})
\label{eq56}
\end{eqnarray}
and
\begin{equation}
<vac |{\cal M}({\bf x}')\phi_i^{cr}({\bf z})\phi_i^{cr}({\bf z})
{\cal M}({\bf x})^\dagger | vac>=g_M({\bf x}'-{\bf x})
<vac | B({\bf x}')\phi_i^{cr}({\bf z})\phi_i^{cr}({\bf z}) | vac>.
\label{eq57}
\end{equation}
We get the contribution to energy from the cr-cr term from equation
(\ref{eq51}) as
\begin{equation} 
h_I^{crcr}=\sum_{n}({3a^2\over 2})^n {1\over n!} {h_{I}^{crcr}}^{(n)}
\label{eq58}
\end{equation} 
where
\begin{eqnarray} 
{h_I^{crcr}}^{(n)}&=&{3a\over \pi S_0}
\left({R_M^2R_\pi^2\over \pi^2}\right)^{3/4}a_n^{-3/2} 
\int Q^2dQq^2dq \int_{0}^{1}dx[W(Q^2,q^2,x^2)]^{-1/4}\nonumber\\ 
&&\left({n_N G^2_{\pi N N}\over M_N}\;\exp(-{q^2R_N^2\over 4})+
{n_\Lambda G^2_{\pi\Lambda\Lambda}\over M_\Lambda}\; \exp(-{q^2R_{\Lambda}^2
\over 4})\right) \nonumber\\ &&\exp\left[-({R_N^2\over 16a_n}+{R_M^2 \over 2})
q^2-{R_\pi^2 Q^2\over 2}\right],
\label{eq59}
\end{eqnarray}
and 
\[[W(Q^{2},q^{2},x^2)]=[(\mu^2+Q^2+{1\over 4}q^2)^2-Q^2q^2x^2].\]
Similarly, the expression for the anhilation-anhilation term of equation 
(\ref{eq51}) can be
calculated. This expression is exactly identical to the above expression
(\ref{eq59}). The energy expectation value for creation-annihilation term
of equation (\ref{eq51}) can be written exactly in the similar way as
\begin{equation} 
h_I^{cran}=\sum_{n}{h_I^{cran}}^{(n)}({3A^2\over 2})^n{1\over n!}
\label{eq62}
\end{equation} 
where
\begin{eqnarray}
{h_I^{cran}}^{(n)}&=&{3a^2~G^2\over 8\pi^4~ S_0}
\left({{4R_M^2R_\pi^2\over \pi a_n(4R_M^2+R_\pi^2)+
R_N^2}}\right)^{3/2} \int Q^2dQq^2dq\int_{0}^{1}dx
[W(Q^2,q^2,x^2)]^{-1/4}\nonumber\\ 
&&\left({n_N\over M_N}\;\exp(-{q^2R_N^2\over 4})+
{n_\Lambda\over M_\Lambda}\;\exp(-{q^2R_{\Lambda}^2\over 4})\right)\nonumber \\
&&\exp\left[-({4R_M^2+ R_\pi^2\over 16}){\bf q}^2-{{4R_M^2R_\pi^2a_{n+1}}Q^2
\over {a_n(4R_M^2+R_\pi^2)+R_N^2}}\right].
\label{eq63}
\end{eqnarray}
Adding equation (\ref{eq58}) and (\ref{eq62}) we have the interaction 
energy as 
\begin{equation} 
h_I=2(h_I^{crcr}+h_I^{cran})
\label{eq64}
\end{equation}

\subsection{Baryon repulsion energy}

We next have to include the energy of repulsion which may arise
from $\omega$-exchange \cite{panda,panda1}. The Hamiltinian is given by the 
simple form
\begin{eqnarray}
H_R&\approx& {g_{\omega NN}^2\over 2m_\omega^2}\int N_\alpha
({\bf z})^\dagger N_\alpha({\bf z})N_\beta({\bf z})^\dagger N_\beta({\bf z})
d{\bf z} 
+2\times {g_{\omega NN}~g_{\omega\Lambda\Lambda}\over 2m_\omega^2}\int N_\alpha
({\bf z})^\dagger N_\alpha({\bf z})\Lambda_\beta({\bf z})^\dagger 
\Lambda_\beta({\bf z}) d{\bf z} \nonumber\\
&+& {g_{\omega \Lambda\Lambda}^2\over 2m_\omega^2}\int \Lambda_\alpha
({\bf z})^\dagger \Lambda_\alpha({\bf z})\Lambda_\beta({\bf z})^\dagger 
\Lambda_\beta({\bf z}) d{\bf z} 
\label{eq65}
\end{eqnarray}
where $m_\omega$, $g_{\omega NN}$ and $g_{\omega \Lambda\Lambda}$
are $\omega$-meson mass, $\omega$ - N coupling and $\omega-\Lambda$
coupling respectively.
Now we calculate the energy due to repulsion through the equation
\begin{equation}
<^AHe({\bf p}')| H_R | ^AHe({\bf p})>= h_R\delta({\bf p}'-{\bf p}).
\label{eq66}
\end{equation}
After a little calculation we get
\begin{eqnarray}
h_R&=&n_N (n_N-1){g^2_{\omega NN}\over{2m_\omega}^2}
\left({1\over 2\pi R_N^2}\right)^{3/2}
+2 n_N n_\Lambda{g_{\omega NN}~g_{\omega\Lambda \Lambda}\over{2m_\omega}^2}
\left({1\over \pi (R_N^2+R_\Lambda^2)}\right)^{3/2}\nonumber\\
&+&n_\Lambda (n_\Lambda-1){g^2_{\omega \Lambda\Lambda}\over{2m_\omega}^2}
\left({1\over 2\pi R_\Lambda^2}\right)^{3/2}.
\label{eq67}
\end{eqnarray}

\subsection{Coulomb repulsion energy}
We next have to include the coulomb repulsion energy. The Hamiltonian
is given by the form
\begin{equation}
H_C=\alpha\int N_{p_{1/2}}({\bf Z}+{{\bf z}\over 2})
^{\dagger}N_{p_{1/2}}({\bf Z}+{{\bf z}\over 2})
N_{p_{-1/2}}({\bf Z}-{{\bf z}\over 2})^{\dagger}N_{p_{-1/2}}
({\bf Z}-{{\bf z}\over 2})\mid z\mid^{-1}d{\bf Z} d{\bf z}
\label{eq68}
\end{equation}
where $\alpha=1/137$.
The coulomb energy is calculated through the equation
\begin{equation}
<^AHe({\bf p}')\mid H_C | ^AHe({\bf p})>=h_C\delta({\bf p}'-{\bf p}).
\label{eq69}
\end{equation}
Using equation (\ref{eq18}) and (\ref{eq68}) we have 
from equation (\ref{eq69}) as
\begin{equation}
h_C={1\over S_0}\alpha({\pi R_N^2})^{-3/2}\int \exp(-{{\bf r}^2\over 2R_N^2})
g_M({\bf r})h_c({\bf r},{\bf z})d{\bf r} d {\bf z},
\label{eq70}
\end{equation} 
where 
\begin{equation}
h_c({\bf r},{\bf z})=\int U^N_\alpha({\bf x}'-{\bf Z}-{{\bf z}\over 2})^*
U^N_\alpha({\bf x}-{\bf Z}-{{\bf z}\over 2})U^N_\alpha({\bf x}'-{\bf Z}
+{{\bf z}\over 2}) U^N_\alpha({\bf x}-{\bf Z}+{{\bf z}\over 2})^*
{1\over |z|}d{\bf Z}
\label{eq71}
\end{equation}
After little algebra, we have
\begin{equation}
h_C=\left({8\over \pi}\right)^{1/2}{\alpha\over R_N}.
\label{eq72}
\end{equation}

\subsection{Meson repulsion energy}
Further, we have taken the pions to be point like,
and assumed that they can approach as near as possible, which is 
physically not correct. If we bring two pions close to
each other there will be an effective force of repulsion because of 
their composite structure. We shall now assume a phenomenological
term corresponding to the meson repulsion as \cite{panda,panda1}
\begin{equation} 
h_M^R=\sum_{n}{h_M^R}^{(n)}({3A^2\over 2})^n{1\over n!},
\label{eq73}
\end{equation}
where 
\begin{equation}
{h_M^R}^{(n)}={3A^2\over S_0}{\left(
{{4R_M^2R_\pi^2}\over{\pi a_n(4R_M^2+R_\pi^2)+R_N^2}}\right)}^{3/2} 0.65
\int\exp{\left[-{{4R_M^2R_\pi^2a_{n+1}}\over
{a_n(4R_M^2+R_\pi^2)+R_N^2}}{\bf q}^2\right]} e^{R_c^2 q^2}d {\bf q}
\label{eq74}
\end{equation}
with $R_c^2=1.2$ fm$^2$.

We have now completed the frame work for dressing of hypernuclei with
scalar and iso-scalar pion pairs. We have also discussed the baryon 
repulsion, meson repulsion and coulomb repulsion. With all these contributions, the situation for nuclear structure can be realistic.
We next minimise the energy given by 
\begin{equation}
E=h_N+h_M+h_I+h_R+h_C+h_M^R
\label{eq81}
\end{equation}
with the parameters $a,~R_M,~R_\pi ,~ R_N$.

\section{RESULTS AND DISCUSSIONS}

In our study we calculate the binding energy using a nonperturbative method.
The nucleon wavefunction and the function for pion dressing are
determined by minimising the energy E given by (\ref{eq81}).
The masses of different particles and the coupling constants used in 
the calculation are given in
Table 1. The pion-nucleon and $\omega$-nucleon coupling constants are 
taken from ref. \cite{panda}. Following Greiner \cite{grein}, we have taken
$\omega-\Lambda$ coupling constant to be $2 \over 3$ of $\omega -N$
coupling constant. This relation has been obtained from SU(6).
The pion-hyperon coupling constant is taken to be 14. 
The  $\pi-\Sigma$ hyperon and $\omega-\Sigma$ hyperon coupling
constants are taken to be the same as those of $\pi - \Lambda$
and $\omega-\Lambda$.

In our calculation, we include the binding energy of $^4$He
by extremising the total energy E given by eq. (\ref{eq81}) as a
check for our programe. The
binding energy for this nucleus is found to be --28.96 MeV
which is the same as earlier value \cite{panda1} and agrees quite well with 
experiment.
The variational parameters $a, ~R_M, ~R_\pi, ~R_N$
obtained from energy minimization are given in Table 2.
Then we added hyperons to this nucleus and repeat the calculation.
For $^{4+n\Lambda }$He and $^{4+n\Sigma}$He, the parameters
$a, ~R_M, ~R_\pi, ~R_N$ are taken same as that for $^{4}$He.
This is because $a, ~R_M, ~R_\pi$ corresponds to meson space
and hence are not likely to change. Again since the number of 
nucleons remain same, $R_N$ is also kept fixed for $^{4+n\Lambda}$He
and $^{4+n\Sigma}$He. In the calculation of $^{4+n\Lambda}$He,
$R_\Lambda$ is the only parameter which is obtained from energy
minimization. Similarly for $^{4+n\Sigma}$He, $R_\Sigma$
is the only variational parameter which is determined from
the minimization of energy. We may suspect that the presence of hyperon will 
change the earlier parameters, but a free variation of all the parameters like
$a$, $R_M$, $R_\pi$, $R_N$ and $R_\Lambda$ or $R_\Sigma$ in each case in the
energy minimisation has negligible effect on the final result. 
The different variational parameters are given in Table 2. 
An analysis of Table 2 indicates that the variational parameter $R_\Lambda$
or $R_\Sigma$ increases as we go to double hypernuclei. This shows that
there developes a $\Lambda$- or $\Sigma$-halo  around the nucleus.
Greiner \cite{grein} obtained a similar result for multi-hyper nuclei with 
their RMF calculation. Since there exists hyperon halos around the
nucleus, it is likely that the hyperons will dissociate under 
peripheral collisions. This may have observational effects.

We then performed a similar
calculation for $^3$He and determined its binding energy by
extremizing the total energy E given by equation (\ref{eq81}). From our
calculation, we find the binding energy for this nucleus to be
--7.09 MeV which agrees quite well with experimental value.
The corresponding variational parameters $a, ~R_M, ~R_\pi,
~R_N$ are given in Table 2. The values of these parameters
are kept constant for multi-hyperon systems based on $^{3}$He
like $^{3+n\Lambda}$He and $^{3+n\Sigma}$He. For $^{3+n\Lambda}$He,
the only variational parameter is $R_\Lambda$ and $R_\Sigma$ is
the only variational parameter for $^{3+n\Sigma}$He. For these
multihypernuclei also, we observe $\Lambda$-halos and $\Sigma$-
halos as in the previous case.

The binding energies of different multi-lambda hypernuclei were
calculated by Greiner \cite{grein}. Their results
agree quite well with our calculated results for $^{4+n\Lambda}$He. We also find
that the binding energies increase when $\Lambda$-hyperons are
added to normal nuclei. However for $\Sigma$-hyperons, the increase
in binding energy is quite small.

Then we have calculated the hyperon separation
energies for different nuclei.
The $\Lambda$ and $\Sigma$ separation energies are given as
\[ B_\Lambda=B(_\Lambda A)-B(A-1)\]
\[ B_\Sigma=B(_\Sigma A)-B(A-1)\]
Here $_{\Lambda /\Sigma}A$ corresponds to a nucleus with $(A-1)$ nucleons and
a hyperon. 
Similarly the $\Lambda\Lambda$ and $\Sigma\Sigma$ separation energies are 
given by
\[ B_{\Lambda\Lambda}=B(_{\Lambda\Lambda}A)-B(A-2)\]
\[ B_{\Sigma\Sigma}=B(_{\Sigma\Sigma}A)-B(A-2)\]
where $_{\Lambda\Lambda /\Sigma\Sigma}A$ stands for total number of nucleons
and hyperons. The separation energies are given in Table 3.

From Table 3, we find that the hyperon separation energy
is reasonably well reproduced for $^{4+\Lambda}$He and $^{3+\Sigma}$He.
However, the separation energy for $^{3+\Lambda}$He is off
from experiment by a factor of 2. The experimental data for double
hyperon separation energy exists only in $^{4+2\Lambda}$He.
The experimental value is 10.6 MeV. From our calculation, we
find this value to be 3.5 MeV. Greiner \cite{grein} has also obtained
a similar result for this nucleus. However, they have observed
that one should include additional hyperon-hyperon interaction
generated by the exchange of mesons with hidden strangeness. 
They have correctly reproduced the double hyperon
separation energy within this frame-work. We also plan to
carry out a similar calculation for the case of multi-hyperon nuclei
by including strange mesons.

\section{CONCLUSION}

We have developed a nonperturbative technique to study the
binding energies of $s$-state hyper nuclei. This model
corresponds to the quantum picture of classical fields
of Walecka in which $\sigma$-meson effects are simulated
through coherent state of pion pairs. Here all the
parameters are calculated variationally through energy
minimization.

We have calculated the binding energies of $^4$He and $^3$He by minimisation
of energy. The agreement with experiment is quite satisfactory. The
binding energies of the $s$-state hypernuclei are calculated by
energy minimisation with respect to $R_\Lambda$ or $R_\Sigma$. The pion
and nucleon parameters are not varied but taken same as those for the
neighbouring nucleus. As lambda-hyperons are added to the nucleus, the
binding energies of the resulting hypernuclei are found to increase.
However, the increase in binding energy is quite small when $\Sigma$ hyperons
are added. We find that hypernuclei develop $\Lambda$- and $\Sigma$-halos.
Similar results are also obtained by Greiner from his RMF calculation. 
Since $\Lambda$- or $\Sigma$- halos develop in these nuclei, the
hyperons may dissociate under peripheral collisions which may have 
observationaleffects. We have calculated the
hyperon separation energies. The agreeement with experiment for hyperon
separation energy is reasonable. The double hyperon separation 
energies are underestimated by a factor around of three. Greiner has
suggested that one should use additional interaction to reproduce the
correct double hyperon separation energy.

Thus without using a nuclear potential and without any
adjustable parameters, we have
been able to reproduce satisfactorily the binding energies
of $^{3}$He, $^{4}$He and the corresponding $s$-state
hypernuclei.

\acknowledgements
The authors are thankful to S.P. Misra for many useful discussions.
The authors are also grateful to Director, Institute of
Physics for providing working facilities and financial support.
One of us (RS) is also thankful to Department of Science and Technology, 
Government of India for financial support.

\appendix
\section{}

We normalize the state through
\begin{equation}
<^AHe({\bf p}')| ^AHe({\bf p})>=\delta({\bf p}'-{\bf p})
\label{eq19}
\end{equation}
which will determine the normalization
constant $N_{R}$ as shown below. Let us define
\begin{equation}
g_{N\Lambda}({\bf x}'-{\bf x})=
<vac |{\cal N}_S({\bf x}') {\cal N}_S({\bf x})^{\dagger} |vac>.
\label{eq20}
\end{equation}
We have anticipated that the above expression is a function
of $({\bf x}'-{\bf x})$. In fact, a direct evaluation of yields
\begin{eqnarray}
g_{N\Lambda}({\bf x}'-{\bf x}) = \prod^{n_N}_{\alpha=1}
\int U^N_\alpha({\bf x}'-{\bf x}_\alpha)^*
U^N_\alpha({\bf x}-{\bf x}_\alpha)d{\bf x}_\alpha
\prod^{n_\Lambda}_{\beta=1}
\int U^\Lambda_\beta({\bf x}'-{\bf x}_\beta)^*
U^\Lambda_\beta({\bf x}-{\bf x}_\beta)d{\bf x}_\beta.
\label{eq21}
\end{eqnarray}
It will be useful to define
\begin{equation}
\rho^N_\alpha({\bf x}'-{\bf x})=\int U^N_\alpha({\bf x}'-{\bf x}_\alpha)^*
U^N_\alpha({\bf x}-{\bf x}_\alpha)d{\bf x}_\alpha,
\label{eq22}
\end{equation}
and
\begin{equation}
\rho^\Lambda_\alpha({\bf x}'-{\bf x})=\int U^\Lambda_\alpha
({\bf x}'-{\bf x}_\alpha)^*
U^\Lambda_\alpha({\bf x}-{\bf x}_\alpha)d{\bf x}_\alpha.
\label{eq23}
\end{equation}
Therefore we have
\begin{equation}
g_{N\Lambda}({\bf x}'-{\bf x})=\prod^{n_N}_{\alpha=1}
\rho^N_\alpha({\bf x}'-{\bf x})
\prod^{n_\Lambda}_{\beta=1}\rho^\Lambda_\beta({\bf x}'-{\bf x})\equiv
g_N({\bf x}'-{\bf x})g_\Lambda({\bf x}'-{\bf x}).
\label{eq24}
\end{equation}
Similarly we define
\begin{equation}
g_M({\bf x}'-{\bf x})=<vac| {\cal M}({\bf x}'){\cal M}({\bf x})^\dagger | vac>.
\label{eq25}
\end{equation}
The above equation is evaluated in the ``mean field aproximation". In fact,
\begin{equation} 
<vac |{\cal M}({\bf x}'){\cal M}({\bf x})^{\dagger}
\mid vac>=\sum^{\infty}_{n=0}{1\over {(n!)^2}}
<vac\mid {(B({\bf x}'))}^n{(B({\bf x})^\dagger)}^n | vac>. 
\label{eq28}
\end{equation} 
In the above we aproximate
\begin{equation} 
<vac| {(B({\bf x}'))}^n{(B({\bf x})^\dagger)}^n | vac> 
=n!\left[<vac| B({\bf x}^{'})B({\bf x})^\dagger | vac>\right]^n
\label{eq29}
\end{equation} 
i.e. we replace products of $B({\bf x}')B({\bf x})^{\dagger}$
by their vacuum expectation values, given as 
\begin{eqnarray} 
&&<vac | B({\bf x}')B({\bf x})^\dagger | vac>\nonumber\\ 
&=&{3\over 2}\int f_M\left({\bf x}'-{{\bf z}_1+{\bf z}_2\over 2}\right)^*
f_M\left({\bf x}-{{\bf z}_1+{\bf z}_2\over 2}\right)f({\bf z}_1-
{\bf z}_2)^*f({\bf z}_1-{\bf z}_2)d{\bf z}_1d{\bf z}_2. 
\label{eq30}
\end{eqnarray}
We then obtain from 
equation (\ref{eq19}), (\ref{eq20}) and (\ref{eq25}) that
\begin{equation}
<^AHe({\bf p}')| ^AHe({\bf p})>  ={(2\pi)}^{-3}
N_R^2{(\pi R_N^2)}^{-3/2}\int g_{N\Lambda}({\bf x}'-{\bf x})
g_M({\bf x}'-{\bf x})e^{(-i{\bf p}'.{\bf x}'
+i{\bf p}.{\bf x})}d{\bf x}'d{\bf x}.
\label{eq26}
\end{equation}
Substituting $({\bf x}'+{\bf x})/ 2={\bf X}$
and $({\bf x}'-{\bf x})={\bf r}$ and comparing with equation (\ref{eq19})
we then obtain from (\ref{eq26}) that, for ${\bf p}=0$,
\begin{equation}
N_R^2{(\pi R_N^2)}^{-3/2}\int g_{N\Lambda}({\bf r})g_M({\bf r})d{\bf r}=1,
\label{eq27}
\end{equation}
We obtain from equation (\ref{eq13}), (\ref{eq14}), (\ref{eq15})
and (\ref{eq16}) that
\begin{mathletters}
\begin{equation}
g_N({\bf r})=\exp{\left (-{n_N{\bf r}^2\over 4R_N^2}\right)}
\label{eq31}
\end{equation}
\begin{equation}
g_\Lambda({\bf r})=\exp{\left(-{n_\Lambda{\bf r}^2\over 4R_\Lambda^2}\right)}
\label{eq32}
\end{equation}
and
\begin{equation}
g_M({\bf r})=\exp\left({3\over 2}a^2\exp\left(-{{\bf r}^2\over 
4R_M^2}\right)\right).
\label{eq33}
\end{equation}
\end{mathletters}
Hence, by equation (\ref{eq27}), $N_R^{-2}$ depends nonlinearly on the 
parameter $a,R_N$ and $R_M$ and infact we explicitly have
\begin{equation}
N_R^{-2}={(\pi R_N^2)}^{-3/2}\sum_{n=0}^{\infty}
\left({3a^2\over 2}\right)^n{1\over n!}\int \exp{\left[-{n_N{\bf r}^2\over
4R_N^2}-{n_\Lambda{\bf r}^2\over 4R^2_\Lambda}
-{n{\bf r}^2\over 4R_M^2}\right]}d{\bf r}.
\label{eq34}
\end{equation}

\newpage
\centerline{\bf Table 1}
\vspace{0.5in}
\begin{tabular}{|c|c|c|c|c|c|c|c|c|}
\hline
$M_N$ &$ M_\Lambda$ & $M_\Sigma$& $\mu$ & $m_\omega$&
$G^2_{\pi N N}/4\pi$ &$ g_{\omega N N}$& $G^2_{\pi\Lambda\Lambda}/4\pi$& 
$g_{\omega\Lambda\Lambda}$  \\ 
\hline    
940&1115.6&1192.6&140&780& 14.6&4.2&14.0&(2/3)$g_{\omega N N}$\\
\hline
\end{tabular}
\newpage
\centerline{\bf Table 2}
\vspace{0.5in}
\begin{tabular}{|c|c|c|c|c|c|c|c|}
\hline
Nucleus & Binding Energy& $a$ & $R_M$ & $R_\pi$ & $R_N$&$R_\Lambda$
&$R_\Sigma$ \\ 
& MeV & & fm & fm & fm & fm&fm \\ 
        \hline 
$^4$He& --28.96 & --0.329& 1.151& 1.696&1.64&--&-- \\
\hline
$^{4+\Lambda}$He& --31.42 & "& "& "&"&2.258&-- \\
\hline
$^{4+2\Lambda}$He& --32.42 & "& "& "&"&2.426&-- \\
\hline
$^{4+\Sigma}$He& --29.92 & "& "& "&"&--&2.605 \\
\hline
$^{4+2\Sigma}$He& --29.84 & "& "& "&"&--&2.87 \\
\hline
$^3$He& --7.09 & "& "& "&1.207&--&-- \\
\hline
$^{3+\Lambda}$He& --12.30 & "& "& "&"&2.282&-- \\
\hline
$^{3+2\Lambda}$He& --15.8 & "& "& "&"&2.4&-- \\
\hline
$^{3+\Sigma}$He& --10.61 & "& "& "&"&--&2.459 \\
\hline
$^{3+2\Sigma}$He& --12.85 & "& "& "&"&--&2.598 \\
\hline
\end{tabular}
\newpage
\centerline{\bf Table 3}
\vspace{0.5in}
\begin{tabular}{|c|c|c|}
\hline
Nucleus & Hyperon Separation Energy&Experimental Value \\
& MeV&MeV \\
\hline 
\multicolumn{3}{|c|}{$B_\Lambda$}\\ \hline
$^{4+\Lambda}$He& 2.46&3.12 \\
\hline
$^{3+\Lambda}$He& 5.21&2.39 \\
\hline
\multicolumn{3}{|c|}{$B_\Sigma$}\\ \hline
$^{4+\Sigma}$He& 0.96& \\
\hline
$^{3+\Sigma}$He& 3.52&3.2 \\
\hline
\multicolumn{3}{|c|}{$B_{\Lambda\Lambda}$}\\ \hline
$^{4+2\Lambda}$He& 3.46&10.6 \\
\hline
$^{3+2\Lambda}$He& 8.71& \\
\hline
\multicolumn{3}{|c|}{$B_{\Sigma\Sigma}$}\\ \hline
$^{4+2\Sigma}$He& 0.88& \\
\hline
$^{3+2\Sigma}$He& 5.76& \\
\hline
\end{tabular}
\newpage
\centerline{\bf Table Captions}

\vspace*{2cm}

\noindent {\bf Table 1.} Masses of nucleon $(M_N)$, $\Lambda$-hyperon 
$(M_\Lambda)$
$\Sigma$-hyperon $(M_\Sigma)$ pion $(\mu)$ and $\omega$-meson $(m_\omega)$ 
in MeV. Pion-nucleon, omega-nucleon, pion-lambda and omega-lambda coupling
constants used in the calculation are also given.\hfil\break

\noindent {\bf Table 2.} Binding energies and variational parameters obtained
from energy minimisation are given for different hypernuclei.\hfil\break

\noindent {\bf Table 3.} The single hyperon separation energy and double
hyperon separation energy for different hypernuclei. The experimental
data are taken from \cite{gibson} and \cite{yama}.\hfil\break
\end{document}